# USING GRID FILES
# FOR A RELATIONAL DATABASE MANAGEMENT SYSTEM

S.M. Joshi, S. Sanyal, S. Banerjee, S. Srikumar

Speech and Digital Systems Group
Tata Institute of Fundamental Research
Homi Bhabha Road, Bombay-400005, India

# USING GRID FILES
# FOR A RELATIONAL DATABASE MANAGEMENT SYSTEM

S.M. Joshi†, S. Sanyal, S. Banerjee, S. Srikumar
satish.m.joshi@gmail.com, sanyals@gmail.com,
suchisrababanerjee@gmail.com
Speech and Digital Systems Group
Tata Institute of Fundamental Research
Homi Bhabha Road, Bombay 400005, India

## ABSTRACT

This paper describes our experience with using Grid files as the main storage organization for a relational database management system. We primarily focus on the following two aspects. (i) Strategies for implementing grid files efficiently. (ii) Methods for efficiency evaluating queries posed to a database organised using grid files.

## KEY WORDS

Relational databases, access methods, grid files, query evaluation

## 1. INTRODUC TION

Since the introduction of the relational model by Codd [1] a large number of database management systems have been implemented using the relational model [2], [3], [4]. While a multitude of ingenious schemes for query evaluation have been devised and used [5], [6], [7], [8], [10], almost all the database management systems have used traditional file structures that provide associative access on a single attribute of the relation, thus making it necessary for the system to maintain many auxiliary access paths on the same relation to facilitate access to tuples based on the value of any one of several attributes or a combination thereof. The burden of maintaining many data structure corresponding to the different access paths, (e.g. A tuple deletion necessitates the updating of all access paths on that relation), is significant particularly for systems meant for small personal machines.

Secondly each of the access paths individually helps in narrowing down the appropriate zone of the physically stored relation that must be searched based on the constraints on only one of the attributes. However such constraints on many attributes together cannot be used meaningfully to further reduce the number of tuples that must be retrieved to satisfy the query.

† Contact Author

A data structure such as the Grid files [12] which provides multi-key access to the records in the file and treats all the keys symmetrically can alleviate both of these problems. With this view a relational data base management system called GREL was designed and is under implementation at the SDS Group, T.I.F.R. which uses grid files as the primary storage mechanism. In this paper we mainly focus on two aspects of this design and implementation.

(1) Several parameters of grid files are left unspecified in the original report viz.

a) Organization of linear scales.
b) Organization of the grid directory.

We discuss alternative strategies for their implementation, the rationale behind them and their impact on the performance of the overall system. Nievergelt et al [12] have already provided experimental data on various performance issues such as average bucket occupancy, estimation of directory size, evaluation of merging polices. Therefore we do not discuss them here.

(2) We also discuss strategies for evaluation of queries on relations which are stored as grid files. A familiarity with the terminology of the relational data model at the level of [13], [14] is assumed.

## 2. A SURVEY OF THE GRID FILE ORGANISATION

Each relation in the database having k attributes defines a hyperparallelpiped (because each attribute may take only a finite number of values) in a k-dimensional space. Each tuple that belongs to the relation is a point in k-dimensional space enclosed in the hyperparallelpiped. Let the relational space be $S= S_1 \text{ x } S_2 \text{ x } S_3 .. \text{ x } S_k$ where each $S_i$ corresponds to the $i^{th}$ attribute of the relation.

On S a partition $P=P_1 \text{ x } P_2 \text{ x } \dots \text{ x } P_k$ is defined which by imposing intervals on each $S_i$ divides the relation space into smaller hyperparallelpipeds which are called grid blocks. Each $P_i$ called a linear scale is an ordered list of values belonging to the domain $D_i$ of the $i^{th}$ attribute. The first value in $P_i$ is the minimum value of $D_i$ and the last is the maximum value in $D_i$. A k-dimensional dynamic array called grid directory is maintained whose elements are in 1:1 correspondence with the grid blocks defined by the partition. Each element of the grid directory contains a pointer to a bucket (which is a storage unit of disk space) that contains tuples which belong to the corresponding grid block. More than one pointer in the grid directory can point to the same bucket.

To fetch a tuple t = ($t_1$, $t_2$... $t_k$) first the grid block in which the tuple lies is detected using the partition P. Next the grid directory element which corresponds to the identified grid block is fetched. Subsequently the tuple is retrieved from the bucket pointed to by the grid directory element. If the grid directory is stored on the disk (which would normally be the case) the 'FETCH' operation requires only two disk accesses.

To insert a tuple t= ($t_1$, $t_2$... $t_k$) in the file a grid block GB which encloses the point ($t_1$, $t_2$... $t_k$) is located and the appropriate bucket is fetched as above. The tuple t is next inserted in the bucket. If the bucket overflows then a new bucket B' is included in the file by refining the partition P if necessary. The partition is refined by splitting that interval which is the projection of GB on some axis $S_i$ which effectively creates $\prod_{\substack{j=1 \\ j \neq i}}^{k} (|Pj| - 1)$ new grid blocks.

Thus to maintain a 1:1 correspondence between the grid blocks and the grid directory elements, it is necessary to insert a new row (which is a k-1 dimensional array) in the $i^{th}$ dimension of the grid directory. To delete a tuple t = ($t_1$, $t_2$... $t_k$) from the file the proper bucket B is located as above and t is deleted from it. If B becomes empty (or the occupancy of B goes below a threshold) the grid block GB that contains t is merged if possible with its neighbors. This may necessitate the deletion of a row from the grid directory.

## 3. AN OVERVIEW OF GREL

GREL is a relational database management system which is to be installed finally on a Motorola 68000 based Exormac system [15]. GREL is written entirely in standard Pascal (with the exception of one module) and was developed on a DEC System-10 where it is currently being tested. GREL is meant for interactive usage through the query language Sequel [16]. Currently it supports a subset of Sequel consisting of (i) all the query statements excluding nested query blocks and set operations, (ii) all the data manipulation facilities, (iii) CREATE and DROP statements. Database views and authorization mechanisms are being implemented.

The major considerations which influenced the design of GREL were the following.

(i) It must be possible to install GREL on small systems similar to Exormac, typically consisting of a 16 bit microprocessor, around 200k bytes of memory and a hard disk drive. This led us to the use of GRID files as the primary storage structure in GREL. It was also the reason for leaving out (for the time being) concurrency and some of the interesting features of Sequel, like integrity assertions and triggers.

(ii) GREL must not implement its own filing system. It must build the necessary access paths over and above the facilities provided by the standard filing system of the host machine.

(iii) It must be easy to transport GREL from one system to another.

Following are the major components of GREL.

(i) Module DISKIO: This is the only module in GREL written in Bliss and comprising of the machine dependent part of GREL (approx 200 lines, 0.5% of the total code). This module receives commands from the rest of the system to acquire and release or to read and write data buckets (a unit of disk storage) which are passed on to the host operating system after appropriate translation.

(ii) Module GRID: This module implements and maintains a grid file for every relation in the database where both user tables and GREL catalogs are treated uniformly as relations.

(iii) Module G-INT: The main job of G-INT is to narrow down the portion of the database that must be searched to satisfy a user query and to search that portion intelligently by making calls on GRID so as to minimize the cost of disk I/O.

(iv) Module PARSER: This module parses Sequel statements that constitute a user query using a standard recursive descent algorithm and outputs a binary tree that represents the parsed Sequel statement. This tree is then interpreted by G-INT.

4. IMPLEMENTATION OF GRID FILES.

The key questions to be discussed in the implementation of grid files are

i) Organization of the grid directory.
ii) Organization of linear scales for each dimension of the grid.

## 4.1 ORGANISATION OF THE GRID DIRECTORY

As described in section 2, the grid directory behaves like a k-dimensional array (for a k-attribute relation) with respect to the 'retrieve' operation. However occasional splitting of grid blocks due to insertions requires that (k-1) dimensional arrays must be inserted in the directory in arbitrary places. The conventional array organization scheme of allocating the array elements in consecutive locations (in say the row major order) implies that the entire grid directory must be rewritten for every split. Since the directory is stored on a disk such a complete rewrite operation is likely to be very costly. Linked lists which are a prime candidate for such dynamic data structure are ruled out because the specifications of the grid directory in [12] stipulate that the address of the desired grid directory element must be computed without making any further disk accesses.

We now describe a scheme [18] for the organization of the grid directory which avoids rewriting the entire directory for every split and which is no more costly than the conventional contiguous allocation scheme as far as accessing a specific grid directory element is concerned. Let the grid directory be an array G of k dimensions. Let $P_1$, $P_2$ ….$P_k$ be the linear scales on the k dimensions. The total number of elements in G is $\prod_{i=1}^{k} ( |Pj| - 1)$. Each Pi is a list of elements $p_{ij}$ consisting of 3 fields. a) A value $V_{ij}$ belonging to the domain of the $i^{th}$ attribute. b) A disk bucket address $A_{ij}$. c) A time stamp $T_{ij}$. (We shall show later that the time stamp need not be stored explicitly).

Suppose a grid block GB is split due to an insert operation by refining the partition on the $i^{th}$ dimension. Let the interval [$V_{ij}$… $V_{i(j+1)}$] be the projection of GB on the dimension i. Then $P_i$ is refined by adding a new entry p* with value V* between $V_{ij}$ and $V_{i(j+1)}$. The time stamp T* is any value larger than any time stamps currently in any of the lists $P_1$, $P_2$, …, $P_k$. Addition of P* to $P_i$ means a new k-1 dimensional array of directory elements (an extra row in the $i^{th}$ dimension of the grid directory) corresponding to the grid blocks whose projection on the $i^{th}$ dimension is the interval [V*….$V_{i(j+1)}$] is to be added to the grid directory. This array is written on the disk and its starting address is put in the address field A* of P*. Note that the old directory is not touched.

Now suppose a tuple is to be retrieved from the grid file. Then the grid block [$V_{1j}$…$V_{1(j+1)}$] x [$V_{2j}$…$V_{2(j+1)}$] x [$V_{kj}$…$V_{k(j+1)}$] in which the tuple falls, is identified by searching the linear scales and the address of the directory element that corresponds to this grid block is computed as follows. The timestamps $T_{ij}$ are compared for 1<=i<=k.

Let $T_{1j}$ be the maximum of these timestamps. Then the required element is located in a k-1 dimensional array having all but the $1^{th}$ dimension of the original k dimensions and having a starting address $A_{1j}$. To compute the address of the desired directory element, we must know the number of entries in each of the lists $P_i$ ($i \neq 1$) at the time when the k-1 dimensional array was written. This is simply found by ignoring all the elements in each $P_i$ with the timestamps larger than $T_{1j}$. Then the desired directory element is located using standard techniques for the calculation of the address of an array element.

It is clear from the above discussion that a refinement of the partition on some dimension does not necessitate a complete rewriting of the grid directory. The grid directory that existed at the time of split is left untouched and a new segment is appended to it. However, accessing a particular directory element to retrieve a specified tuple does seem to be more complicated. Specifically the conventional array organization for the grid directory requires that the linear scales be searched only once to identify the grid block that encloses the tuple t whereas our scheme (as described above) requires that the linear scales be scanned twice. In the first place the linear scales are all assumed to be small enough to be in the main memory so the cost of scanning them is not very much. Secondly we shall show that by organizing the linear scales cleverly, the second scan can be easily eliminated.

## 4.2 ORGANISATION OF THE LINEAR SCALES

As described in section 3, each element of the linear scale consists of three fields, a value V belonging to the domain of the corresponding attribute of the relation, a disk address and a timestamp. A basic assumption made in [12] is that the linear scales are small enough to be accommodated in the core. To satisfy this assumption we must try to reduce the size of the elements in the linear scales.

First we shall eliminate the timestamp. This is done by maintaining a composite sequential list called PARTLIST of all the elements of the linear scales in the order in time in which the elements are added to the linear scales. Each element of P* of PARTLIST has three fields, V* and A* as before and a third field D* where the value of D* is i if P* belongs to the linear scale for the $i^{th}$ attribute. Note that the length of the D* field can be much smaller than the timestamp. Each time a data bucket overflows causing a grid block to split and requiring a refinement of the scale on some attribute i, an element (V*, A*, i) is appended at the tail of the PARTLIST with V*, A* computed as in Section 4.1 i.e. the timestamp of an element is represented implicitly by its position in the list.

To retrieve a tuple $t = <t_1, t_2, \ldots, t_k>$ PARTLIST is searched and elements $P_{i1}$, $P_{i2}$ for $1<=i<=k$ of PARTLIST are located with the property that $D_{i1} = D_{i2} = i$ and $V_{i1}<=t_i<=V_{i2}$ for $1<=i<=k$ and there do not exist $p'_{i1}$, $p'_{i2}$ such that $V_{i1}<V'_{i1}<=ti<V'_{i2}<V_{i2}$ for $1<=i<=k$. Suppose for some j, $P_{j1}$ occurs last in PARTLIST among all $P_{i1}$, $1<=i<=k$ then $A_{j1}$ is the starting address of the k-1 dimensional array consisting of all the k dimensions except for the $j^{th}$, which contains an element pointing to the bucket that contains t. The position of the directory element within the array is computed by keeping track of: (i). The number of elements (V, A, D) of PARTLIST such that D= i and $V < V_{i1}$ and $i \neq j$. (ii) The number of elements (V, A, D) of PARTLIST such that D=i and $i \neq j$ that occur before $P_{j1}$ in PARTLIST. This procedure requires only one scan of PARTLIST.

To reduce the size of elements of PARTLIST further we use a prefix coding (such as that used in [17]) for the value field of the elements. This is necessary because the query language Sequel supported by GREL allows attributes whose domains are character strings of arbitrary lengths. In practical applications it is not uncommon to have character strings of length fifty or more (e.g. Titles of books, addresses of persons etc.) To store the entire character string in the value field of elements of PARTLIST is often wasteful.

Therefore initially when a new element is added to PARTLIST only the first (most significant) word of the attribute value is stored in the value field (less than one word is not attempted due to the complexity of implementation). This will lead to a failure in trying to refine an interval on some attribute when many tuples are inserted which are too close to each other to be differentiated based only on the most significant word. In such a case the value field of the new element to be added to PARTLIST is extended as follows.

Note that the only attributes whose domains are character strings can have lengths more than one word and the ordering assumed on such domains is the lexicographic ordering. Now suppose a bucket overflow necessitates the refinement of the linear scale on some attribute i, by splitting an interval $[V_1..V_2]$. Then there exist elements (V', A', i) and (V'', A'', i) in PARTLIST such that the first $w_1$ words of $V_1$ are identical to V' and the first $w_2$ words of $V_2$ are identical to V''. If $w_1 < w_2$ then V' otherwise V'' is extended to a length max $(w_1, w_2)$ by padding it with words containing the minimum representable integer.

A value V* is chosen such that $V' < V^* < V''$ and the tuples in the bucket that overflowed can be divided into sets such that the $i^{th}$ attribute value of tuples in one set (the first max ($w_1$, $w_2$) words of it) is less than V* and for the other set it is greater than or equal to V*. (V* could simply be the word by word average of the first max ($w_1$, $w_2$) words of the attribute values). If no such V* is found then the same procedure is repeated to find a value of length max ($w_1$, $w_2$) + 1 and so on till the maximum allowed length of the $i^{th}$ attribute is reached. In that case the linear scales on the other attributes are considered for refinement. Otherwise the new element (V*, A*, i) with A* computed appropriately is added to PARTLIST without any modification to either V’ or V’’.

The schemes are outlined above to serve two purposes. They reduce the size of the elements of PARTLIST and eliminate an extra scan of PARTLIST while retrieving a tuple.

We close this section by posing a question which we shall try to answer in the next section. Efficient processing of range queries was one of the major objectives of grid files. Whereas in our organization of the grid directory we seem to have destroyed the natural correspondence between the value ordering of the attribute domains on one hand and the elements of the directory and linear scales on the other. Have we impaired the ability to process range queries efficiently in doing so?

## 5. EVALUATION OF QUERIES

We view a query as Boolean valued expression e ($R_1$, $R_2$...$R_n$) where $R_1$, $R_2$... $R_n$ are the relations over which the query is defined. The expression is built from operands (such that $R_i.A_j$ denoting attribute $A_j$ of relation $R_i$ and literal values) and operators such as =, ≠, $<$, $>$, $<=$, $>=$, NOT, AND, OR which have the usual meanings. The query expression is represented as a binary tree, a common technique used in many compilers.

The process of evaluating such a query can be split into two major steps. (i) Identification of as small a portion of physically stored relations as possible such that all the tuples that can possibly satisfy the query can be found in this portion. (ii) Efficient retrieval of all the tuples in the identified region.

Algebraic transformation [6], use of ‘sargable predicates’ [7], decomposition of a query into irreducible components [5] are some of the examples of the first step. Selection of the least cost access paths [7], [11], the various join algorithms [9], [10], variable selections in decomposition [5] are some of the examples of the second step.

Algebraic manipulation of [6] or query decomposition in [5] (apart from the cost formulae used in variable selection) is independent of any specific storage structures and thus can be used over and above the schemes we described here. The access path selection methods such as [7], [11] assume the existence of specific data structures and cannot be used directly for grid files simply because for a given relation there are no access paths to choose from. However some of the techniques in [7] for selecting the proper join method and the use of 'interesting orders' to do so can be adapted to be used in [7]. However we do not discuss these issues here.

## 5.1 RESTRICTING THE SCANS ON INDIVIDUAL RELATIONS

Each query on relations $R_1$, $R_2$...$R_n$ defines a subregion of the hyperparallelpiped of each $R_i$ such that all and only those tuples of $R_i$ which get mapped into points enclosed in this subregion satisfy the query. Our aim is to identify this subregion as precisely as possible. To simply the presentation the following discussion will be in terms of a query on only one relation R with three attributes $A_1$, $A_2$, $A_3$ whose domains are integer ranges O...$M_1$, O...$M_2$, O...$M_3$ respectively.

### BASIC QUERY EXPRESSIONS

Basic query expressions are of the form (i) R.$A_i$ op C and (ii) R.$A_i$ op R.$A_j$ where op is comparison operator and C is a constant in the range O...Mi, 1 <=1, j <=3. For each comparison operator the subregion defined by (i) and (ii) are shown in Figs. 1 and 2.

Regions defined by (i) are always parallelepipeds and can be represented simply by their projections on the dimensions of the 3-dimensional space of R. i.e. A region $C_1$ is represented by $<I_{11}, I_{12}, I_{13}>$ where each $I_{1j}$ is an interval $[L_{1j}...U_{1j}]$ on the $j^{th}$ axis. We shall use this notation through out. However the regions defined by (ii) are not so easy to represent. Also in our opinion meaningful queries of the form (ii) will arise rarely in real applications. (Such queries do arise in database of geometric objects but Sequel is hardly the appropriate query text books [13], [14] nor the various papers that deal with query evaluation give a single example of such queries. Therefore we represent such regions also by their projections on the dimensions of the space of R. e.g. the parallelepiped made by a dotted line in Fig. 2.

### PROCESSING THE AND OPERATOR

We consider query expressions of the from e = expl AND exp2 where each of exp1, exp2, will in general be represented by a set of non overlapping parallelepipeds $H_1$, $H_2$ respectively (because of the OR operator explained below). Thus the region defined by e is represented by a set of parallelepipeds H such that for $C_1$ in $H_1$ and $C_2$ in $H_2$ if $C_1 \cap C_2$ is not empty (i.e. $I_{1j} \cap I_{2j}$ not empty for all j, 1 <=j<=3) then $C_1 \cap C_2$ belongs to H. $C_1 \cap C_2 = < I_{11} \cap I_{21}, I_{12} \cap I_{22}, I_{13} \cap I_{23} >$. An algorithm to compute H from $H_1$, $H_2$ is given in Fig. 4. Note that since all members of $H_1$ (also $H_2$) is not overlapping the members of H will also be non overlapping.

PROCESSING THE OR OPERATOR

We now consider query expressions of the form e ≡ expl OR exp2. As before exp1, exp2 define sets of non overlapping parallelepipeds $H_1$, $H_2$ respectively.

In processing the AND operator we were aided by the distributivity of the set intersection operator over the Cartesian product. In case of the OR operator the region defined by e is the union of the regions defined by exp1 and exp2. However the set union operator does not distribute over the Cartesian product i.e. the union of two parallelepipeds may yield a concave region that cannot be represented simply by its projections on the dimensions of the relation space. Fortunately such regions can always be represented by a set of non overlapping parallelepipeds. The qualification 'non overlapping' is important to ensure that no tuple in the region defined by the query will be fetched twice when these parallelepipeds are used subsequently to restrict the scan on the relation R. e.g. Consider two parallelepipeds $C_1$ and $C_2$. Let H be the set of parallelepipeds that represent the union of $C_1$ and $C_2$. The set H can be computed by the algorithm in Fig. 5.

Of course the implementation is a bit more complex because the cardinality of H depends on whether $C_1$ or $C_2$ is split and on the choice of j in Fig. 5 e.g. It is better to choose a j where $((L_{2j} < L_{3j})$ AND $(U_{2j} = U_{3j}))$ OR $((L_{2j} = L_{3j})$ AND $(U_{2j} < U_{3j}))$ holds rather than where $(L_{2j} < L_{3j})$ AND $(U_{2j} < U_{3j}))$ holds. Using this algorithm then the set of parallelepipeds defined by the query e can be computed from $H_1$, $H_2$ as in Fig.6. It must be noted here that the algorithm is not free from the dangers of a combinatorial explosion i.e. The number of non overlapping hyperparallelpipeds in which $C_2$ gets split can be $(2^k - 1)$ for a k-attribute relation in the worst case. Hopefully this will not happen in the normal case.

We do not consider the Boolean NOT operator because it can be removed from the query expression using DeMorgan's rules e.g. NOT ((R. $A_1 < 20$) AND (R.$A_2 = 0$)) = (R. $A_1 >= 20$) OR (R. $A_2 \neq 0$).

The sub-regions of the space of the relation R defined by every query can thus be found as a set of non overlapping parallelepipeds. This set is then passed as a parameter to the procedure that will actually search the grid file for R. Using this set the extent of the search will only be limited to those tuples that fall within the region defined by the query.

This technique is similar to the use of 'sargable predicates' in System-R [7] but more powerful on two counts. First, the use of sargable predicates does not reduce the number of tuples fetched but the number of calls on the search procedure. Secondly, sargable predicates that refer to only one of the attributes can be used meaningfully because in [7] ultimately a single access path defined on only one attribute is chosen to scan the relation.

## 5.2 SCANNING A RELATION

The problem we consider in this section is the following. Given a hyperparallelpiped C we have to locate all the grid directory elements that correspond to grid blocks which have a non empty intersection with C and fetch the data bucket pointed to by the directory element. (The following discussion is also in terms of the relation R described in the last section)

The main considerations in the design of the scanning algorithm are the order in which the directory elements are laid out on the disk (we must scan the grid directory in the same order to reduce disk I/O) and the effects of sharing of a data bucket among many grid blocks. We shall ignore the later for the moment.

Recall that the directory is stored in several separate pieces in the order in time in which bucket overflows caused refinements of the linear scales. The linear scales are also stored in the same order as a composite list called PARTLIST.

Looking in turn at every element p of the PARTLIST we decide whether any of the directory elements contained in the piece of the grid directory pointed to by p, corresponds to a grid block that has a non empty intersection with C.

This is done as follows. Let C= $< L_1 \ldots U_1, L_2 \ldots U_2, L_3 \ldots U_3 >$. Locate elements of PARTLIST, $p_{1i}$, $p_{2i}$ for all i, 1<= i<=3. (Recall that the fields of $p_{1i}$, $p_{2i}$ are referred to as $p_{1i}$.V, $p_{2i}$.A etc.). $p_{1i}$, $p_{2i}$ have the property that $p_{li}$. D = $p_{2i}$. D = i, $p_{1i}$. V <= $L_i$, p2i. V <= $U_i$ and there does not exist any p* in PARTLIST such that p*. D = i and either $p_{1i}$. V < p*. V <= $L_i$ or $p_{2i}$.V < p*. V <= $U_i$. Now if p.D = i and $p_{1i}$.V <= p.V <= $p_{2i}$.V then the piece of the directory pointed to by p.A contains some of the directory elements that we are interested in. Each directory element e in this piece uniquely corresponds to (i.e. the address of e within the piece can be determined by) a pair of elements of PARTLIST p', p" that occurs before p in PARTLIST and p'. D ≠ p''.D ≠ i. Let p'. D = j and p''. D = k. If additionally $p_{1j}$. V <= p'. V<= $p_{2j}$. V and $p_{1k}$. V <= p''. V <= $p_{2k}$. V then we fetch the corresponding directory element e and the data bucket pointed to by e. Notice that the above procedure requires that for every element p of PARTLIST the portion of PARTLIST before p must be scanned once more. However since PARTLIST is small enough to be stored in the main memory, the cost of doing so will not be excessive.

Data buckets can be shared among many grid blocks. Therefore to ensure that a shared bucket is fetched only once we must add some more information to the directory elements. For a k-attribute relation we add an array called SHARED of k Boolean elements to each directory element e. If e. SHARED [i] is true then the bucket pointed to by e is shared between the grid block b that corresponds to e and the grid block next to b in the decreasing direction of the $i^{th}$ dimension of the grid. The array SHARED in a directory element e needs to be updated every time the bucket pointed to by e overflows which can be done with little or no extra cost. This array together with the fact that every set of grid blocks that share a bucket forms a hyperparallelpiped is sufficient to ensure that no shared bucket is fetched twice. When a directory element e is fetched using p, p', p'' as described above, the grid block corresponding to e is bounded below by p.V, p'.V, p''.V in the dimensions p.D, p'.D, p''.D respectively. The bucket pointed to by e is fetched iff for every j in {p.D, p'.D, p''.D} either e.SHARED[j] is false or for every j such that e.SHARED[j] is true, the corresponding V field is equal to $p_{1j}$.V. It is easy to verify that for every set of directory elements that point to the same bucket and that lie in the given region C, the above condition will hold for one and only one member of the set.

So despite having destroyed the correspondence between the value ordering of attribute domains and the directory elements it is possible to process range queries efficiently.

This search procedure will be used both for answering single relation queries and also for performing joins and Cartesian products of relations which we describe in the next section.

## 5.3 JOINS AND CARTESIAN PRODUCTS OF RELATIONS

In this section we consider queries that refer to relations $R_1$, $R_2$…$R_k$, k > 1 and contain terms of the type $R_i$. A op $R_j$. B where op is a comparison operator in addition to the the basic expressions described earlier. Such terms are called join terms and attributes A, B are called the join attributes of $R_i$, $R_j$ respectively.

Two basic methods of computing joins can be found in almost all the implemented systems (i) The nested join and (ii) The merge join [9], [10].

The cost of the nested join (measured as the number of disk accesses made) of k relations $R_1$, $R_2 \ldots R_k$ is $\sum_{i=1}^{k}(\prod_{j=1}^{i} Bj)$ where $B_j$ buckets is the size of $R_j$. The nested join method essentially computes the Cartesian product of relations and filters out those tuples from the product which do not satisfy the join terms in the query.

The merge join method requires that the relations $R_1$, $R_2 \ldots R_k$ be sorted on their join columns. The cost of a merge join is then $\sum_{i=1}^{k}$ ($B_i$ + sortcost ($R_i$)) where sortcost ($R_i$) is the cost of sorting $R_i$. If the relations are already sorted or if the cost of sorting is small then the merge join clearly outperforms the nested join, for example when there is an index on the join column of each relation. However, the following limitations of the merge join must be noted, (i) The merge join can be used meaningfully only for equijoins. It obviously cannot be used for Cartesian products, (ii) It is difficult to use the merge join to join a relation $R_1$ with relations $R_i$ and $R_j$ on two different attributes A and B respectively of $R_1$, since it requires the re-sorting of the result of one of the joins $R_1$ with $R_i$ or $R_1$ with $R_j$ whichever is done first.

Given a query E defined over relations $R_1$, $R_2 \ldots R_k$, for each pair of relations $R_i$, $R_j$ such that there exists a join term in E joining $R_i$, $R_{j;}$ we must select one of the two join methods.

In [7] an elaborate method to do such a selection is given which can be adapted to our environment by changing the cost formulae used. However we decided not to do so primarily to simplify the implementation. Secondly the power of these methods will really be exercised if the queries involve a large number of joins with conflicting requirements. It is difficult to think of applications wherein queries often contain more than 3 relations to be joined simultaneously and a majority of the joins are not equijoins. Also in our case we noted that whenever it is possible to join two relations using the merge join (subject to the constraints given above), with high probability, it will also be profitable to do so. This is because of the following property of the grid files.

For some relation R, Let $p_1$, $p_2$,...$p_m$ be elements of PARTLIST such that $p_i$.D = j for all i, 1<=i<=m and $p_i$. V < $p_{(i+1)}$. V for 1<=i<m. Then by passing appropriate parameters to the search procedure described above the tuples of R can be retrieved in such a way that all the tuples with the $j^{th}$ attribute value in the interval [$p_i$. V... $p_{(i+1)}$. V] are returned before those with the $j^{th}$ attribute value in the interval [$p_{(i+1)}$. V...$p_{(i+2)}$. V] i.e. approximately sorted on the $j^{th}$ attribute of R. Therefore the cost of sorting the entire relation is likely to be small. (Incidentally, this property is also useful to process the ORDER BY and GROUP BY Clauses in Sequel).

So the simple method implemented in GREL is as follows. The set of relations $R_1$, $R_2$,...$R_k$ referred to in a query is divided in to disjoint subsets $S_1$,$S_2$…$S_m$, m<= k as follows.

(i) Each relation belongs to one and only one subset.

If $R_i$, $R_j$ belongs to some subset $S_1$ then the following conditions hold.

(ii) There exists a term $R_i$. A = $R_j$. B in the query for some attributes A and B of $R_i$, $R_j$ respectively.

(iii) There does not exist a term in the query of the form ($R_i$. C op $R_j$. D) where op is a comparison operator other than = or C is not the same as A or D is not the same as B.

(iv) If there is a term that joins $R_i$ or $R_j$ with a relation $R_k$ on attributes other than A or B respectively then $R_k$ does not belong to $S_1$.

(v) There does not exist a term of the form exp1 op exp2 where op is any comparison operator and exp1, exp2 are arithmetic expressions containing attributes of $R_i$ and $R_j$ (in the same term) as operands.

Relations belonging to each subset are then joined by a multi-way merge join and the resulting relations for each group are joined by a nested join.

In retrospect, though this scheme appears too simplistic. Firstly because the conditions (i)... (v) do not force a unique partition on the set of relations. Thus we must try to make an optimal choice. Secondly the notion of interesting orders as used in [7] must be introduced in the criteria used for choosing a partition. In other words grid files do not seem to make the task of joining relations any easier than conventional access methods except that sorting the relation on some attribute is cheaper and the problem of selecting the proper access path does not exist.

## 6. CONCLUDING REMARKS

The major difference between grid files and conventional single attribute access paths is that a grid file replaces a collection of such access paths by a single data structure. Such an amalgamation helps in many ways. It simplifies the problem of maintaining access paths under deletions and insertions and of selecting the proper access path for retrieval. The symmetric treatment of all attributes makes it possible to limit very precisely the extent of a relation scan by simultaneously making use of the constraints on all the attributes specified by a query.

On the other hand due to the symmetric treatment we are likely to loose some precision (No. of tuples that satisfy a query / No. of tuples fetched) in answering queries as compared to the precision in answering the same query when there are conventional access paths on the attributes referred to in the query e.g. a query that specifies a range constraint on a key of the relation. Secondly from a user's point of view all attributes of a relation are not equally important. Thus some of them are referred to in queries more often than the others.

In the context of grid files the above considerations translate to the problems of (i) choosing an appropriate subset of the set of attributes of a relation to define the grid on that relation and of (ii) controlling the granularity of the linear scales on every such chosen attribute to limit or enhance the precision of answering queries that refer to that attribute.

As an example consider some typical results of our experiments. Our test database consisted of two relations, the relation BOOKS with attribute names and types as follows ACNQ (CHAR (5)), TITLE (CHAR (50)), AUTHOR (CHAR (25)), CLASSNO (CHAR (5)), PUBLISHER (CHAR (25)), and YEAR (INTEGER). The second relation SMALLBOOKS was identical to BOOKS except that only TITLE, AUTHOR, and YEAR were used to define the grid. Identical data (taken from S.D.S Group library of TIFR) was inserted in both.

We implemented two different splitting polices and tried them on BOOKS and SMALLBOOKS.

(i) Split on all attributes in a strictly round robin manner.

(ii) Consider each attribute in turn for splitting. If a split at the mid-point of the relevant interval of the linear scale on that attribute succeeds then do so otherwise try the next attribute. If the attempt fails for all attributes then split the relevant interval on any attribute at points other than the mid-point.

Note that the second scheme will lead to a finer partition on that attribute along which the tuples in the database are uniformly distributed.

We were mainly interested in observing the effects of (a) The number of attributes of a relation used for defining the grid and (b) The splitting policy, on the size of the grid directory. For various data sets under both the splitting polices similar bucket occupancy was observed for BOOKS and SMALLBOOKS. However the size of the grid directory and the redundancy in the directory (i.e. the ratio of total number of elements in the directory to number of data buckets) was markedly different for the two relations under the same splitting policy and for the two relations under different splitting policies. Some typical results are given in Table 1. Table 2 gives the number of disk accesses made to answer various queries on BOOKS under the two splitting policies and also the results for the same queries on SMALLBOOKS. In these experiments TITLE and AUTHOR were the two most frequently used attributes. The improvement in the performance of queries on these attributes as a result of increased granularity of the linear scales on them is evident.

Acknowledgements

We are most grateful to Prof. Jurg Nievergelt of ETH for introducing us to grid files and for his subsequent interest in our project. We thank Dr. M. N. Cooper for the encouragement given to us during the course of this project. We are grateful to Mr. R. Prabakaran, TIFR and NCSDCT for providing the facilities for this project.

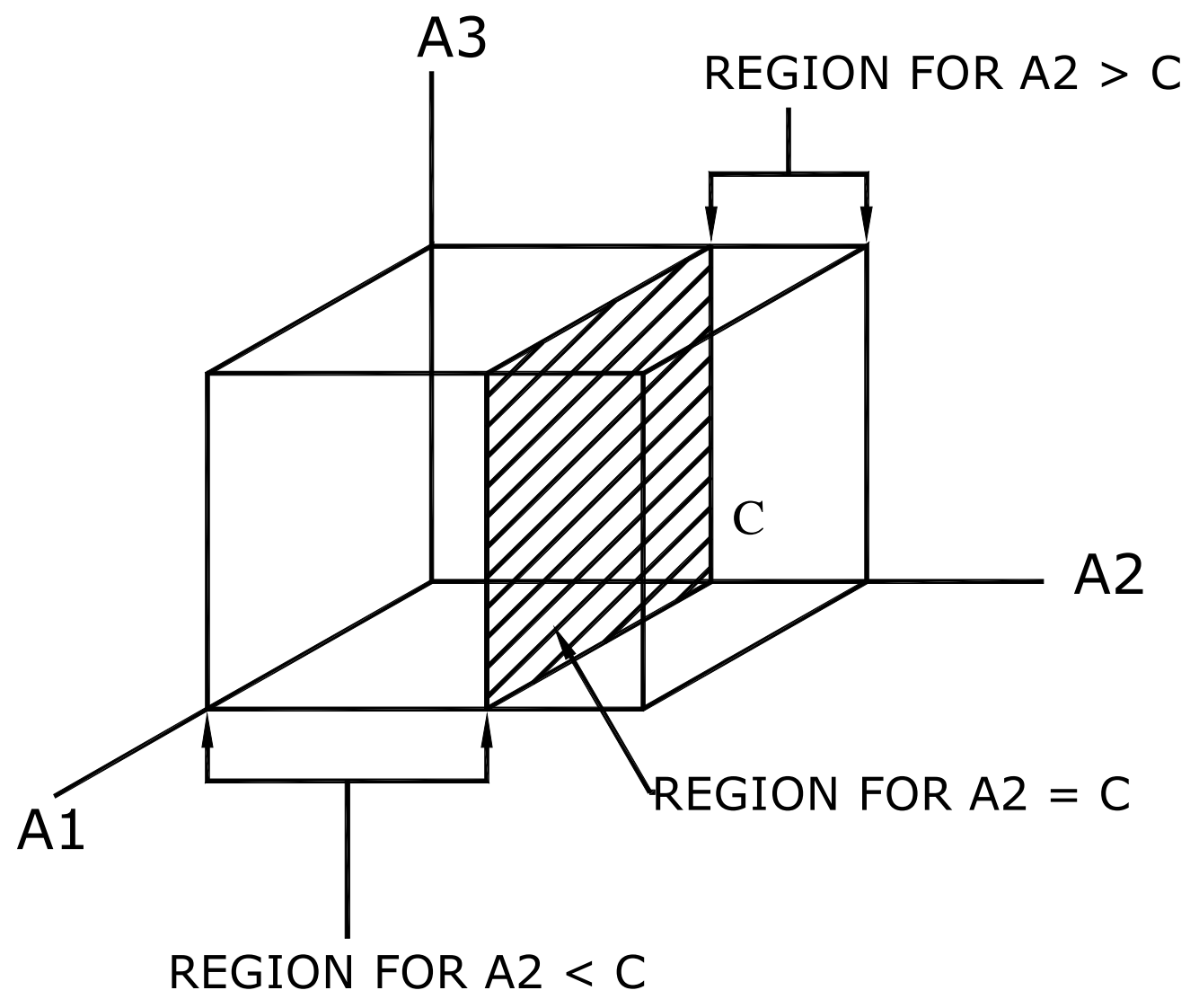
A3
REGION FOR A2 > C
C
A2
A1
REGION FOR A2 = C
REGION FOR A2 < C

FIG - 1

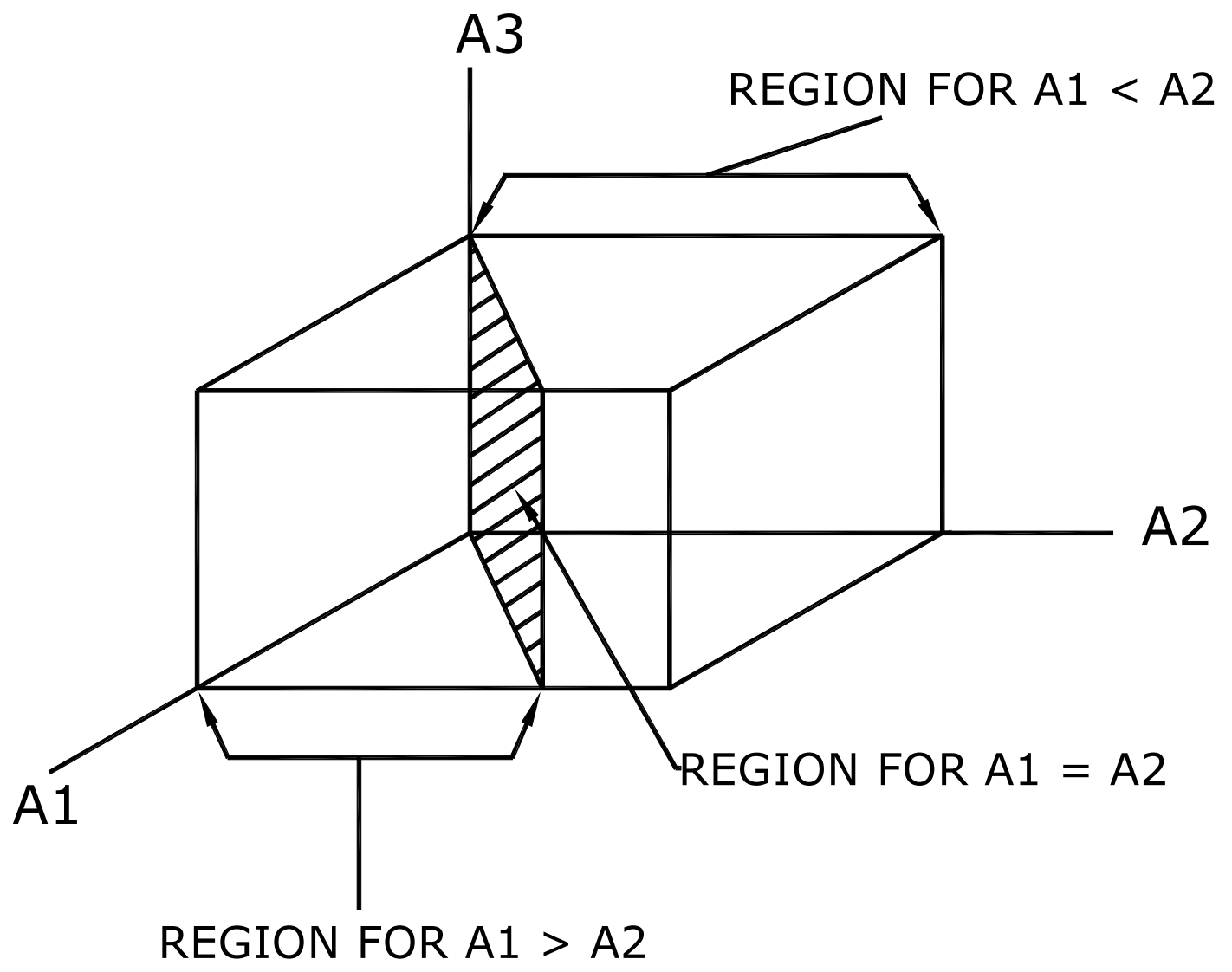
A3
REGION FOR A1 < A2
A2
A1
REGION FOR A1 = A2
REGION FOR A1 > A2

FIG - 2

```
H, H1, H2: sets of parallelepipeds;
begin
   H : = Ø ;
   for all C1 in H1 do
     begin
        for all C2 in H2 do
          begin
             if C1 ∩ C2 ≠ Ø then H : = H ∪ {C1 ∩ C2}
          end
      end
 end
```

Figure 3

Procedure SPLIT ($C_1$, $C_2$: parallelepipeds; return H: set of parallelepipeds);

```
C2, C3, C4: parallelepipeds;
begin
   C3 : = C1 ∩ C2; H : = Ø ;
   if C3 = Ø then H : = {C1, C2}
   elseif C3 = C1 then H : = {C2}
   elseif C3 = C2 then H : = {C1}
   else ( * split C2 * )
     begin
         while C3 ≠ C2 do
           begin
               C4 : = C2;  C5 : = C2;
               Choose j, 1 <=j<3 such that I3j ≠ I2j;
               if L2j < L3j then
                   begin
                       I4j : = [L2j…L3j – 1] ; I5j : = [L3j...U2j] ;
                   end
             else
              begin
                  I5j : = [L2j...U3j] ; I4j : = [U3j + 1 ... U2j] ;
              end;
             H : = H ∪ {C4} ; C2 : = C5 ;
           end ;
         H : = H ∪ {C1}
  end SPLIT.
```

Figure 4

```
         H, H1, H2, H3: sets of parallelepipeds;
         C3 : parallelepiped ;
         begin
            H : = Ø ;
           for all C1 in H1 do
              begin
                 for all C2 in H2 do
                    begin
                       SPLIT (C1, C2, H3);
                       if H2 = {C2} then go to SKIP ;
                       else H2 : = ( H2 ∪ (H3 – {C1}) ) – {C2} ;
                    end ;
                 H : = H ∪ {C1} ;
SKIP:            H1= H1 – {C1}
              end ;
           H : = H ∪ H2
         end.
```

Figure 5

TABLE 1

| RELATION | BUCKET OCCUPANCY | REDUNDANCY |
|---|---|---|
| BOOKS | 70 % | 7.7 |
| SMALLBOOKS | 74 % | 4.2 |

| SPLITTING SCHEME | OCCUPANCY | REDUNDANCY |
|---|---|---|
| (i) | 68 % | 45.2 |
| (ii) | 70 % | 7.7 |

TABLE 2

| ATTRIBUTE | NO. OF PARTITIONS | | |
|---|---|---|---|
| | SPLITTING POLICY | | |
| | ( i ) | ( ii ) | |
| ACNO | 4 | 1 | 1 |
| TITLE | 4 | 5 | 9 |
| CLASSNO | 4 | 2 | 1 |
| AUTHOR | 4 | 6 | 7 |
| PUBLISHER | 2 | 2 | 1 |
| YEAR | 3 | 2 | 2 |
| | BOOKS | | SMALLBOOKS |

| QUERY EXPRESSION | NO. OF DISK ACCESSES | | |
|---|---|---|---|
| | BOOKS | | SMALLBOOKS |
| | SPLITTING POLICY | | |
| | ( i ) | ( ii ) | |
| TITLE = ‘DISTRIBUTED CONTROL’ | 10 | 10 | 5 |
| AUTHOR = ‘ULMAN’ | 10 | 8 | 5 |
| YEAR = 80 | 13 | 29 | 28 |
| ( YEAR = 80 ) and ( TITLE = ‘DISTRIBUTED CONTROL’) | 8 | 7 | 3 |

This Paper [USING GRID FILES FOR A RELATIONAL DATABASE MANAGEMENT SYSTEM] was referred to in the seminal Paper by

J. Nievergelt, Hans Hinterberger, Institute for Informatik, ETH; Kenneth C. Sevcik, University of Toronto. The Grid File: An Adaptable, Symmetric Multikey File Structure ACM Transactions on Database Systems (TODS), Volume 9, Issue 1, March, 1984. Pages: 38-71, ISSN: 0362-5915, as [Reference 12].